# HOW TO PATCH ACTIVE PLASMA AND COLLISIONLESS SHEATH: PRACTICAL GUIDE

Igor D. Kaganovich

*Plasma Physics Laboratory, Princeton University, Princeton, New Jersey, 08543, USA*

Most plasmas have a very thin sheath compared with the plasma dimension. This necessitates separate calculations of the plasma and sheath. The Bohm criterion provides the boundary condition for calculation of plasma profiles. To calculate sheath properties a value of electric field at the plasma-sheath interface has to be specified in addition to the Bohm criterion. The value of the boundary electric field and robust procedure to approximately patch plasma and collisionless sheath with a very good accuracy are reported.

## I. Introduction

The calculation of plasma profiles is nowadays a routine task. In most plasmas employed in applications the Debye length $\lambda_D$ is small compared with the plasma half width $L$. Resolving the small Debye length throughout the whole plasma requires solving Poisson equation, which is challenging computationally because the electric field has to be obtained from small differences between the electron and ion densities. To avoid the inconvenience, the standard procedure is to separate the plasma and sheath regions, and to employ the quasineutrality condition in the plasma region instead of Poisson's equation. The Bohm criterion – setting the ion velocity equal to the ion sound velocity - gives the boundary condition for the plasma region and uniquely defines plasma profiles. In contrast to the plasma region, the Bohm criterion is not sufficient for a unique determination of sheath properties.

If the sheath potential is much larger than the electron temperature, it follows from the Boltzmann relation that the electron density in the sheath can be neglected, the plasma sheath boundary can be assumed infinitely thin, and the electric field at the plasma-sheath interface can be set to zero. This approach has been successfully applied for calculating sheath parameters in dc (Child-Langmuir law [1]) and rf discharges [2, 3, and 4].

The relevant question is: is it possible to calculate sheath properties with higher accuracy? If all regions with a length of order $\lambda_D$ and a potential drop order the electron temperature $T_e$ have to be resolved, an accurate patching between the plasma and the sheath region has to be performed. In general, it requires either a direct numerical solution of Poisson's equation throughout the plasma and sheath regions or applying matched asymptotic approximations, as described in Refs. 5, and 6, and in references there in. Numerical simulation of Poisson's equation for the whole discharge is computationally intensive and inefficient. The utilization of matched asymptotic approximations requires a great deal of mathematical expertise and is not very robust for engineering purposes. Therefore, there were a number of attempts to patch the plasma and sheath approximately.

Poisson's equation is a second order equation, and it requires two boundary conditions: the potential on the wall, and another condition set at the plasma-sheath interface. Because the position of the plasma-sheath interface is unknown a priori, the values of both the potential and the electric field have to be specified. In Refs. 7 and 8 the value $T_e/(e\lambda_{Ds})$ was proposed for the electric field at the plasma-sheath patching point, where $\lambda_{Ds}$ is the Debye length corresponding to the plasma density $n_s$ at the plasma-sheath interface. This electric field has been utilized as the boundary condition to join the plasma and sheath in discrete plasma-sheath models and was used in the calculations of dc [7] and rf [3] sheaths.

This approach has been recently criticized in Ref. 9, where it was claimed that such a procedure results in "the disjunction between the plasma and sheath". In their response [10], the authors of Ref. 8 refuted this claim, and explained that the sheath solution in Ref. 9 was taken with zero electric field at the plasma-sheath boundary instead of $T_e/(e\lambda_{Ds})$.

In this Letter, a new procedure for approximate patching is proposed. A new value for the electric field at the plasma-sheath boundary is determined from numerical calculations and the theory of the transition layer to be $0.962[T_e/(e\lambda_{Ds})](\lambda_{Ds}Z/c_s)^{3/5}$, where $Z$ is the ionization frequency, $c_s=\sqrt{T_e/M}$ is the ion sound speed, and $M$ is the ion mass. This value agrees with the theory of the transition layer between the plasma and sheath [5,6]. In addition, it was found that accounting for the small transition region between the plasma and sheath regions, which has a width of order $(\lambda_{Ds}/L)^{4/5}\lambda_{Ds}$ and a potential drop of order $(\lambda_{Ds}/L)^{2/5}T_e$, yields an approximate solution which is very close to the exact solution. These numerical findings verify the theory of the transition layer described in Refs. 5, 6, and 9.

## II. Basic equations

We shall employ fluid equations in one dimension in the collisionless approximations. The same notation is used as in Ref. 9. These equations consist of the continuity equation

$$\frac{d}{dx}(n_i v_i) = Zn_e, \tag{1}$$

the ion momentum conservation equation

$$M\frac{d}{dx}(n_i v_i^2) = -en_i\frac{dV}{dx}\,, \qquad (2)$$

and the Boltzmann relation governing electrons density

$$n_e = n_0 \exp\left(\frac{eV}{T_e}\right). \qquad (3)$$

Here, the subscripts i and e denote ion and electron quantities, respectively, and the subscript 0 corresponds to the central density values at $x=0$. $V$ is the potential.

The potential is given by Poisson's equation

$$-\frac{d^2V}{dx^2} = 4\pi e(n_i - n_e)\,. \qquad (4)$$

The boundary conditions for the system of Eqs.(1)-(4) are: at the symmetry axis ($x = 0$), $V = 0$, $dV/dx = 0$, $n_e = n_i \equiv n_0$; at the wall ($x = L$), $V = V_w$, where $V_w$ is the wall potential, see Appendix 1. The ionization frequency $Z$ is an eigenvalue of the system of Eqs.(1)-(4).

The system of Eqs.(1)-(4) is known to yield results very close to the exact ion kinetic approach [11, 12]. Because of its simplicity, it has been widely employed in theoretical and engineering studies.

In the limit $\lambda_{Ds} << L$, the potential can be determined from the quasineutrality condition $n_i = n_e$. Substituting the Boltzmann relation Eq. (3) into the quasineutrality condition yields the plasma potential: $V = T_e/e\ln n_i$. Following Ref. 9, and normalizing Eqs. (1)-(3) in the plasma region ($n_e = n_i$) with $N_i = n_i/n_0$, $N_e = n_e/n_0$, $\phi = -eV/T_e$, $U = v_i/c_s$ gives [9]

$$N_i = \frac{1}{1+U^2} = e^{-\phi}\,, \qquad (5)$$

$$\frac{dU}{dx} = \frac{Z}{c_s}\frac{1+U^2}{1-U^2}\,. \qquad (6)$$

Equation (6) has the solution $xZ/c_s = 2\arctan U - U$ [5, 13]. Eq. (5) is singular at the point $U = 1$, meaning that the plasma can not overcome the ion sound velocity in this solution. Bohm showed that sheath can be patched with the plasma only if $v_i \geq c_s$ [14]. Therefore, at the plasma-sheath interface ($x=L_p$) the Bohm criterion $v_i = c_s$ holds. From the Bohm criterion, one readily finds $Z = (\pi/2 - 1)c_s/L_p$, and the plasma solution gives $n_s = n_0/2$ and $V_s = -T_e/e\ln 2$ at the point $U = 1$.

## III. Patching sheath and plasma

The Poisson equation (4) is a second order equation, therefore, it requires two boundary conditions. One is the value of the potential at the wall $V_w$, and another boundary condition is determined from correct patching with the plasma. Using direct numerical integration of the system of Eqs. (1-4) for a wide range of parameters $\lambda_{D0}/L$, where $\lambda_{D0}$ is the Debye length corresponding to the central plasma density $n_0$, it was determined that the value of the electric field at the point where $v_i = c_s$ agrees with the expression

$$E_s = T_e(\lambda_{D0}/L)^{3/5}/(e\lambda_{D0})\,, \qquad (7)$$

to within 10% accuracy and is independent on the wall potential. The results of the simulations are gathered in Table 1.

| | $\lambda_{D0}=\lambda_{Dr}x\sqrt{10}$ | $\lambda_{D0}=\lambda_{Dr}$ | $\lambda_{D0}=\lambda_{Dr}/\sqrt{10}$ | $\lambda_{D0}=\lambda_{Dr}/10$ |
|---|---|---|---|---|
| $(\lambda_{D0}/L)^{3/5}$ | 0.102 | 0.051 | 0.0257 | 0.0130 |
| $\phi_w=1$ | 0.102 | 0.049 | 0.0240 | 0.0119 |
| $\phi_w=5$ | 0.112 | 0.052 | 0.0243 | 0.0121 |
| $\phi_w=10$ | 0.117 | 0.053 | 0.0244 | 0.0121 |

Table 1. The value of normalized electric field $eE\lambda_{D0}/T_e$ for different values of $\lambda_{D0}/L$ and wall potentials. The reference value $\lambda_{Dr}/L=0.7071x10^{-2}$ was taken from Ref.9.

Table 1 lists values of the normalized electric field $eE\lambda_{D0}/T_e$ at the point where $v_i = c_s$ for $L=1$, four different values of $\lambda_{D0}/L$ (in a wide parameter range), and three values of the wall potentials $V_w=-1,5,10T_e$. The reference value $\lambda_{Dr}/L=0.7071x10^{-2}$ was taken to be the same as in Ref. 9. The other values of $\lambda_D/L$ include the value half an order of magnitude larger than the reference value, half an order of and an order of magnitude smaller than the reference value. The second line in Table 1 shows the value of $(\lambda_{D0}/L)^{3/5}$. From Table 1, it is clearly seen that all the values in a given column are close to each other, meaning that the value of the normalized electric field $eE\lambda_{D0}/T_e$ at the point where $v_i = c_s$ is close to the value given by Eq.(7) and is independent of the wall potential.

Knowing the value of the electric field at the plasma-sheath interface, the sheath properties can be determined. Neglecting the increase in the ion flux due to ionization in the bulk of the sheath region enables one to readily integrate Eqs.(1,2), giving

$$n_i = \frac{\Gamma_s}{c_s\left[1 + \dfrac{2e(V_s - V)}{T_e}\right]^{1/2}}\,, \qquad (8)$$

where $V_s$ is the potential at the plasma-sheath interface and $\Gamma_s$ is the ion flux in the sheath. Substituting the ion density Eq.(8) and electron density Eq.(3) into Poisson's equation and integrating once gives:

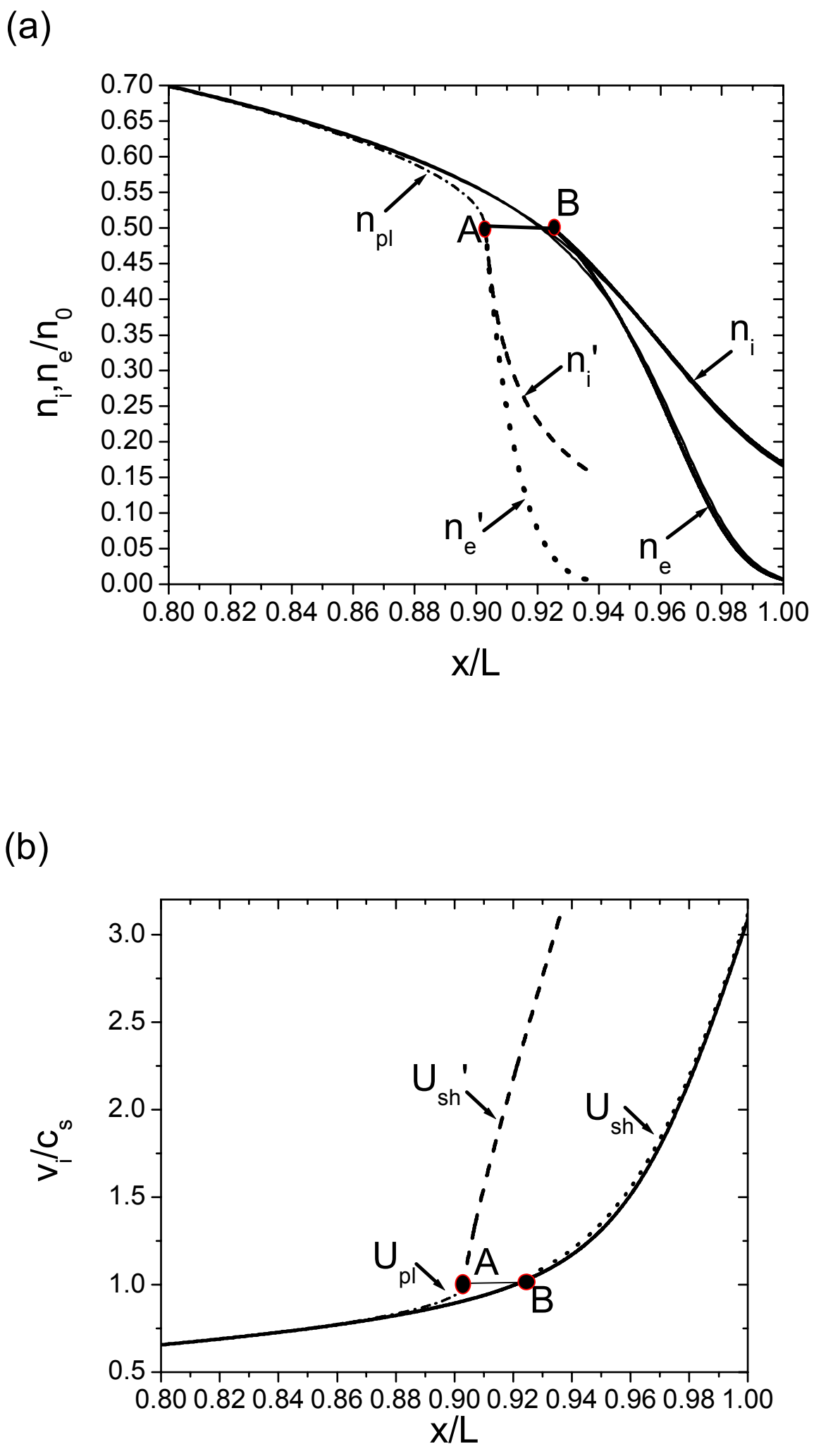

Fig.1(a) Ion and electron density profiles and Fig. 1(b) Ion flow velocity profiles calculated from the full system of equations (1)-(4) (solid lines), and approximate solutions in the sheath using Eq. (9) (dashed line for ion density and dotted line for electron density). Approximate solutions in the sheath with the electric field at the plasma-sheath boundary given by Eq.(7) and the location shifted from the point A (x=0.907) to the point B (x=0.921) are practically indistinguishable from the exact solution. Prime denotes the sheath solution with the electric field at the plasma-sheath boundary $E_s = T_e/(e\lambda_{Ds})$, as proposed in Ref.7. The plasma solution of Eqs.(5), and (6) is shown with the dash-dotted lines. The discharge conditions are the same as in Fig.1 of Ref.9 : $\lambda_{D0}/L=0.7071x10^{-2}$, $V_w=-5T_e/e$.

$$\frac{1}{2}\left(\frac{d\eta}{dy}\right)^2 = \frac{1}{2}\left(\frac{d\eta}{dy}\right)_s^2 + \gamma(1+2\eta)^{1/2} + \exp(-\eta) - (1+\gamma) \quad , \tag{9}$$

where following the same notation as in Ref. 9, theses normalized quantities were introduced: $y = (x - x_s)/\lambda_{Ds}$, $\eta = e(V - V_s)/T_e$, $d\eta/dy\,|_s = \lambda_{Ds} e E_s / L T_e$ and $\gamma = 2\Gamma_s/n_0 c_s$. $\gamma$ is larger than unity, and accounts for the additional ionization in the transition layer and adjacent sheath region, (see appendix II for details). Equation (9) is readily integrated, yielding ion and electron density profiles in the sheath, as shown in Fig.1(a).

Fig.1 shows very good agreement between the exact and approximate sheath solutions, in contrast to the claim of Ref.9. In Ref. 9, zero boundary electric field at the plasma sheath interface was used, thus, producing an oversimplified patching, as described in Ref.10.

Fig.2 depicts the electric field as a function of the normalized potential (*-eV/T*$_e$). This figure is similar to Fig.2 of Ref.9 but instead of patching the plasma solution Eq.(6) and the sheath solution Eq.(9) using $E_s = 0$, $E_s$ given by Eq. (7) was used. Apparently, such a patching of plasma and sheath solutions yields an electric field profile, which is very close to the exact solution, in disagreement with the claim of Ref. 9.

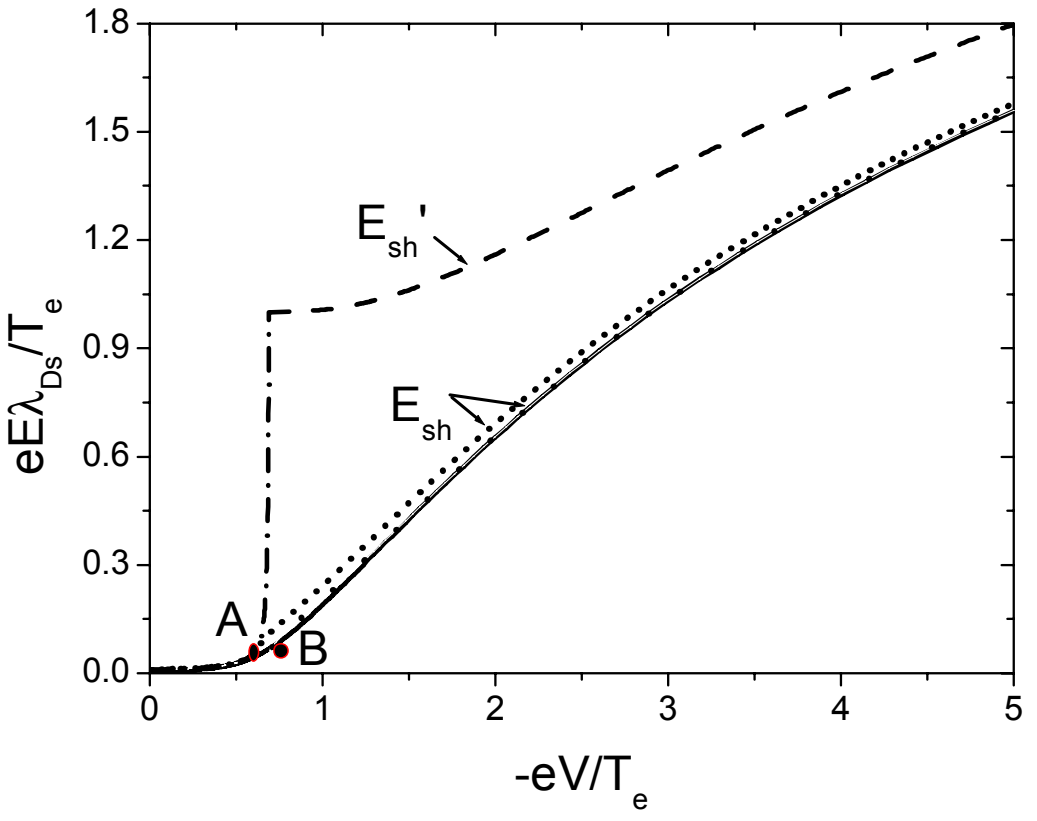

Fig.2 The variation in the electric field in units of $T_e/\lambda_{Ds}$ as a function of the potential in units of ($T_e/e$). The conditions are the same as in Fig.1. The finely dotted line corresponds to the sheath solution patched at the point where the plasma solution given by Eqs. (5) and (6) has the same electric field $E_s$ as given by Eq. (7), corresponds to point A ($-eV/T_e = 0.623$). The coarsely dotted line (practically indistinguishable from the exact solution) patches the sheath solution at the point B ($-eV/T_e = 0.709$) with the patching position being shifted by $0.085$. The shifting distance corresponds to $1/2\delta\phi_{tr} = 0.085$, where $\delta\phi_{tr}$ is given by Eq.(20). The plasma solution of Eqs.(5) and (6) is showed with dash-dotted lines. The dashed line corresponds to the sheath solution utilizing the value of the electric field at the plasma-sheath boundary $E_s = T_e/(e\lambda_{Ds})$, as proposed in Ref 7.

The patching of the sheath solution of Eq. (9) with the plasma solution Eqs. (5) and (6) at point where $E=E_s$ apparently gives continuous of electric field profiles, because the electric field is assumed continous in the patching. This disagrees with the claim of Ref.9 [10]. At the same time, in accord with Ref.9 the value of the electric field $E_s = T_e/(e\lambda_{Ds})$ (proposed for patching in Ref.7) corresponds to the point of exact solution $V=-3T_e$ which is far inside the sheath, namely at $x=0.985$ ,. Thus, it neglects part of the sheath (from *x=0.907* to *x=0.985*) and correspondingly neglects the potential difference (*3-0.62*) $T_e$, as compared with the exact solution.

From Fig.2, it is obvious that even patching using the value of electric field $E_s$ in Eq. (7) does leave out a part of the exact solution. Shifting the patching point by a distance in potential $\delta V_{tr} = (2Z\lambda_{Ds} / c_s)^{2/5} T_e / (2e)$ produces very good agreement between the approximate sheath solution Eq.(9) and the exact solution. This "disjunction" between the plasma and sheath clearly indicates the necessity of a special transition layer between plasma and sheath.

## IV. Transition layer

The transition layer appears due to a sonic singularity in plasma equations. As shown by Bohm [14], the sheath electric field can be smoothly patched with the small electric field in the plasma (small compared with the sheath) only if the ion flow velocity at the plasma-sheath boundary is larger or equal to the ion sound velocity. Therefore, a transition through the ion sound velocity should occur in the plasma. It follows from Eqs.(6) that the ion sound velocity can not be exceeded in plasma with a slab geometry, and, therefore, the ion sound velocity must be reached at the boundary between the sheath and plasma regions. The situation is different for non slab geometry. If plasma expands in some kind of plasma nozzle with cross-sectional area *A(x)*. The continuity equations become instead of Eqs.(1) and (2)

$$\frac{d}{dx}(n_i v_i A) = Z n_e A \tag{10}$$

$$M\frac{d}{dx}(A n_i v_i^2) = -eAn_i\frac{dV}{dx}, \tag{11}$$

and Eq. (6) describing plasma region is modified to

$$(1-U^2)\frac{dU}{dx} = \frac{Z}{c_s}(1+U^2) - U\left(E_1 + \frac{1}{A}\frac{dA}{dx}\right), \tag{12}$$

where the electric field $E_1 = E - (-T_e d\ln N_i / dx)$ is the difference between the actual electric field and the electric field obtained with the quasineutrality assumption and the Boltzmann relation. From Eq.(12), it is obvious that a transition from subsonic to supersonic flow is possible only if a plasma channel is expanding, for example in cylindrical or spherical geometries. A necessary condition for avoiding the sonic singularity is that right hand of Eq. (12) equals zero where $U = 1$. In slab geometry it is only possible if $E_1 > 0$. Transition through sonic speed in a slab geometry requires $E > (-T_e d\ln N_i / dx)$, i.e., breaking quasineutrality.

Correspondingly, to obtain the mathematical structure of the transition layer one has to solve Poisson 's equation near the sonic point. In the sonic point ($x = x_s$, $v_i = c_s$), $n_i \cong n_0 / 2$. Series expansion of the ion flux gives

$$\Gamma_i \cong n_0 / 2[c_s + Z(x - x_s)], \tag{13}$$

and the ion velocity becomes [making use of Eq.(2)]

$$M(v_i^2 - c_S^2)/2 = -e(V - V_s) - Z(x - x_s), \tag{14}$$

yielding the space charge near the sonic point

$$n_i - n_e = \frac{n_0}{2}\left[\frac{1 + Z / c_s(x - x_s)}{\sqrt{1 + 2\phi - 2Z / c_s(x - x_s)}} - e^{-\phi}\right]. \tag{15}$$

Expanding the space charge in Eq.(15) to the first nonzero term in $\phi$ and $x - x_s$ gives the Poisson equation near the sonic point

$$\frac{d^2\phi}{dx^2} = \lambda_{D0}^{-2}\left[\frac{1}{2}(\phi - \phi_s)^2 + \frac{Z}{c_s}(x - x_s)\right]. \tag{16}$$

The same result can be obtained by differentiating the Poisson equation (4) and substituting the ion and electron density derivatives from Eqs.(1)-(3), which readily gives [5]

$$\frac{d^3\phi}{dx^3} = \lambda_{D0}^{-2}\left[\frac{d\phi}{dx}\left(e^{-\phi} - \frac{N_i}{U^2}\right) + \frac{2Ze^{-\phi}}{c_s U}\right]. \tag{17}$$

Equation (17) is exact and describes both plasma and sheath regions. In the limit $\lambda_{D0} << c_s / Z$, the electric field can be determined by setting the right hand side of Eq. (17) is to zero. This procedure fails at certain $\phi = \phi_s$ where $N_i = U^2 e^{-\phi}$. At this point ($x = x_s$), the ion velocity is close to the ion sound velocity $U \approx 1$, because the quasineutrality condition $N_i = e^{-\phi}$ holds in the nearest vicinity of this point. In the neighborhood of $\phi = \phi_s$, the left hand side of Eq. (17) must be also accounted for. Performing Taylor expansion near $\phi = \phi_s$: $e^{-\phi} - N_i / U^2 = (U^2 e^{-\phi} - N_i)/U^2 \approx 1/2[U^2(\phi) - 1] \approx (\phi - \phi_s)$, Eq. (17) becomes

$$\frac{d^3\phi}{dx^3} = \lambda_{D0}^{-2}\left[\frac{d\phi}{dx}(\phi - \phi_s) + \frac{Z}{c_s}\right]. \tag{18}$$

Integrating Eq.(18) yields Eq.(16).

Equation (16) is a nonlinear, nonhomogeneous differential equation. The scaling of the solution for potential $\delta\phi_{tr}$ and transition layer width $\delta x_{tr}$ can be estimated from Eq. (16), looking for a solution in the form $\phi - \phi_s = \phi_{tr} F(x / x_{tr})$, where $F = O(1)$. Near the point $x = x_s$, all terms of Eq. (16) should be of the same order, therefore

$$\frac{\delta\phi_{tr}}{(\delta x_{tr})^2} = \frac{1}{2}\lambda_{D0}^{-2}(\delta\phi_{tr})^2, \ \frac{1}{2}(\delta\phi_{tr})^2 = \frac{Z}{c_s}\delta x_{tr}. \tag{19}$$

The solution of Eq.(19) is

$$\delta\phi_{tr} = \left(\frac{2\lambda_{Ds}Z}{c_s}\right)^{2/5}, \ \delta x_{tr} = \lambda_{Ds}\left(\frac{c_s}{2\lambda_{Ds}Z}\right)^{1/5}. \tag{20}$$

This scaling Eq.(20) was received in matched solutions in Ref. 5, 6 and 13. Note that it is necessary to account for the ionization term (last term in Eq. (18),(16) in order to receive a smooth matching of the plasma and sheath solutions [15]. The function $F(\xi = x / x_{tr})$ is obtained from the equation

$$\frac{d^2F}{d\xi^2} = F^2 + \xi \,. \tag{21}$$

The boundary condition corresponds to the quasineutral region at $\xi < 0$ $F = -\sqrt{-\xi_0}, dF / d\xi = 1/2\sqrt{-\xi_0}$ where $-\xi_0 >> 1$ is any large number.

The plot of function $F(\xi)$ is shown in Fig.3.

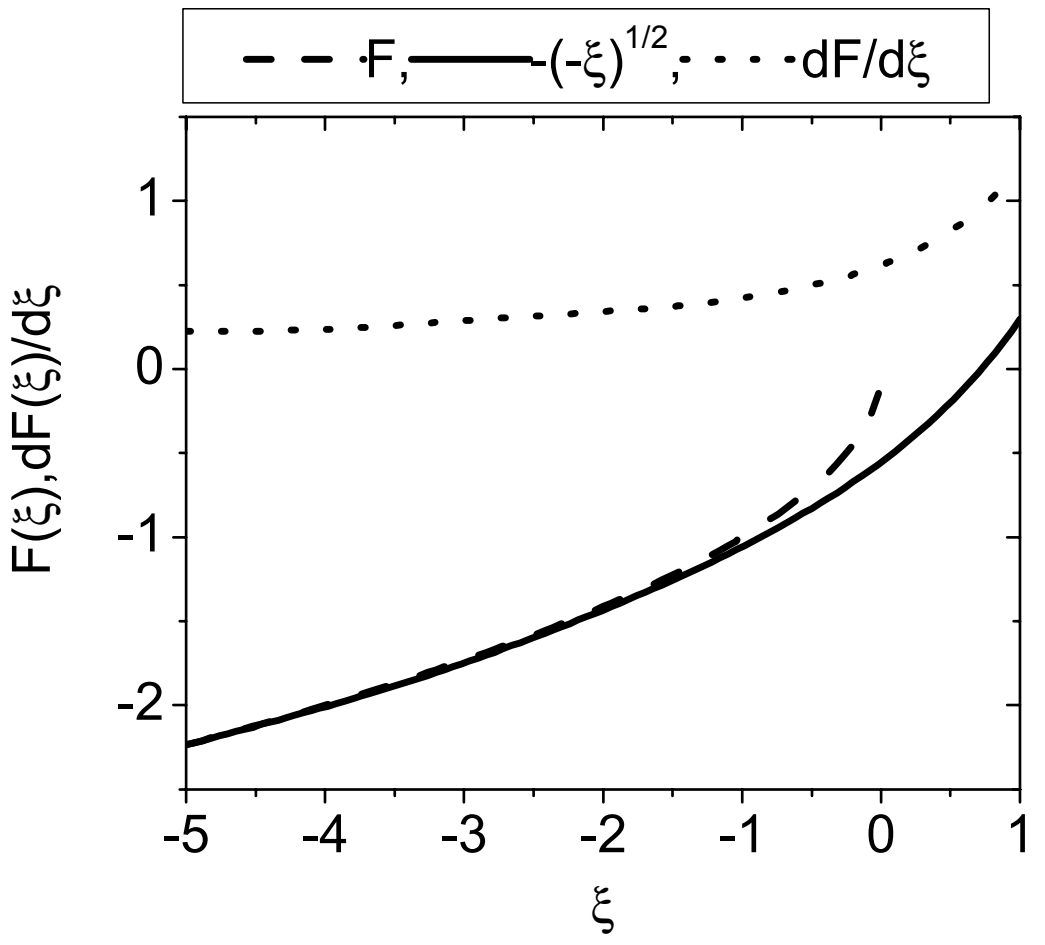

Fig.3 The plot of normalized potential and electric field in the transition layer $\phi - \phi_s = \phi_{tr} F(\xi = x / x_{tr})$ and the normalized electric field $E = (\phi_{tr} / x_{tr}) dF / d\xi$ .

In Fig.3, one can see that the function $F(\xi)$ breaks the quasineutral solution $F_{pl}(\xi) = -\sqrt{-\xi}$ at $\xi > -1$. The sonic point corresponds to *F(y)=0* (see Eq.(14), second term on the right hand side is small compared with the first term). At this point, $dF / d\xi = 0.962$ and the value of electric field is therefore $E_s = 0.962\, \delta\phi_{tr} / \delta x_{tr}$. Substituting scales for $\delta\phi_{tr}$ and $\delta x_{tr}$ from Eqs.(20) gives

$$E_s = 0.962 \frac{T_e}{e\lambda_{Ds}} \left( \frac{2\lambda_{Ds} Z}{c_s} \right)^{3/5} . \tag{22}$$

Substituting $Z = (\pi/2 - 1) c_s / L$ for collisionless plasma gives the same value for the electric field at the sonic point as Eq.(7) but with a factor $0.962 [\sqrt{2}(\pi - 2)]^{3/5} / \sqrt{2} = 0.907$. As can be seen from Table.1, the value of electric field in Eq.(7) reduced by a factor 0.907 agrees better with the numerical simulation results at small Debye lengths (see the last two columns).

To summarize, the transition region is a distinct region, which can not be attributed to either the sheath or plasma regions. Indeed, though in this region the quasineutrality condition approximately holds (see Fig.1 $x \approx 0.90 \div 0.94$), the electric field can not be determined from the quasineutrality condition. (see Eq.(16)). From the other side, even though Poisson's equation is used to determine the properties of the transition region, this region is not a sheath if the Bohm concept of the sheath is used: a *"region, characterized by negligible electron density"* [14].

## V. Conclusion

An approximate procedure to patch sheath and plasma is proposed. The sheath and plasma are patched at the point where the value of the electric field $E_s = 0.962\, T_e / e\lambda_{Ds} (2\lambda_{Ds} Z / c_s)^{3/5}$, the transition layer is accounted simply by shifting the sheath solution from the patching point by a distance $\delta x_{tr} = \lambda_{Ds} / (2\lambda_{Ds} Z / c_s)^{1/5}$ and the potential by $\delta V = -(2Z\lambda_{Ds} / c_s)^{2/5} T_e / (2e)$. For most practical purposes, the value of $\delta V << T_e$ is very small compared with sheath potential and can be neglected.

## ACKNOWLEDGEMENTS

The author is grateful to Ron Davidson, Raoul Franklin, Valery Godyak, Kyle Morrison, Yevgeny Raitses, Edward Startsev, and Gennady Shvets for helpful discussions. This research was supported by Department of Energy via the University Research Support Program of Princeton Plasma Physics Laboratory.

## Appendix I. Note on wall potential

The wall potential is to be determined by equating the ion and electron fluxes. The ion flux is $\Gamma_i \cong n_s c_s$, from Eq.(13). The electron flux is given by an integral over the electron velocity distribution function (EVDF) for all electrons with velocity directed toward the wall

$$\Gamma_{ew} = n_{ew} \left( \frac{m_e}{2\pi T_e} \right)^{1/2} \int_0^\infty v_x \exp\left( -\frac{m_e v_x^2}{2T_e} \right) dv_x \,. \tag{A.I.1}$$

Integrating yields

$$\Gamma_{ew} = n_{ew} \left( \frac{T_e}{2\pi m_e} \right)^{1/2} , \tag{A.I.2}$$

where the electron density at the wall $n_{ew}$ is to be determined from the Boltzmann relation $n_w = n_s \exp[e(V_s - V_w)/T_e]$. Therefore, equating the ion and electron fluxes at the wall gives

$$V_s - V_w = \frac{T_e}{2e} \ln\left( \frac{M}{2\pi m_e} \right). \tag{A.I.3}$$

Equation (AI.3) is correct for a collisionless sheath and either collisional or collisionless plasmas. For the case of a collisionless plasma, $V_s = -T_e / e \ln 2$.

Though in the present paper only the Boltzmann relation is used, it is necessary to note that the Boltzmann relation is not accurate for electrons leaving the plasma and being lost at the wall (so called loss cone). The Boltzmann relation requires a Maxwellian EVDF and that the electrons are trapped in a potential well. Because of fast losses to the wall, the EVDF is non-Maxwellian in the loss cone. Therefore, it is necessary to solve the kinetic equation for fast electrons to obtain a correct EVDF in the loss cone, and, subsequently, to predict the wall potential. Examples of such a calculation are given in Ref.16. The analytical solution of the EVDF in the loss cone is given in Ref.17.

**Appendix II. Ionization in the sheath region**

Ionization in the sheath region is determined by the integral

$$I_{sh} = Z\int_{x_s}^{x_w} n_e(x)dx \, . \qquad \text{(A.II.1)}$$

Changing the variable of integration from *x* to the normalized potential $\eta = e(V - V_s)/T_e$, the integral (A.II.2) becomes

$$I_{sh} = Zn_s\lambda_{DS}\int_0^{\eta_w} \frac{e^{-\eta}}{d\eta / dy} d\eta \, , \qquad \text{(A.II.2)}$$

where the normalized electric field $d\eta / dy$ is given by Eq. (9). The function $(1+2\eta)^{1/2} + \exp(-\eta) - 2 \approx \eta^3 / 3$ at $\eta \ll 1$, for that reason, the integral (A.II.2) diverges if $E_s = 0$ ($d\eta / dy = \sqrt{2/3}\ \eta^{3/2}$) [15]. Therefore, the main contribution to the integral is at small $\eta$. Numerical integration shows that within 5% accuracy

$$\int_0^{\eta_w} \frac{e^{-\eta}}{d\eta / dy} d\eta \approx \frac{1.3}{\sqrt{E_s\lambda_{Ds} / L}} \qquad \text{(A.II.3)}$$

in the wide range $E_s\lambda_{Ds} / L = 0.01 - 0.3$. Substituting this estimate for the integral Eq. (A.II.3) into Eq. (A.II. 2), one obtains an equation for $\gamma = 2\Gamma_s / n_0 c_s$, where $\Gamma_s = n_0 c_s / 2 + Zn_0 / 2\delta x_{tr} + I_{sh}$. Finally, we obtain

$$\gamma = 1 + \frac{Z\delta x_{tr}}{c_s} + \frac{1.3\lambda_{Ds}}{\sqrt{\lambda_{Ds} e E_s / T_e}} \, . \qquad \text{(A.II.4)}$$